\documentclass[final,5p,times,twocolumn,authoryear]{elsarticle}

\usepackage{amssymb}
\usepackage{amsmath}
\usepackage{lipsum}

\usepackage{mathrsfs}

\journal{International Journal of Geometric Methods in Modern Physics}

\begin{document}

\begin{frontmatter}




\title{Matter-free gravitational collapse and the equivalence principle}
\author{Juri Dimaschko}
\address{Technische Hochschule Lübeck, Mönkhofer Weg 239, 23562 Lübeck, Germany}

\begin{abstract}
We investigate the gravitational collapse of a spherically symmetric degenerate wormhole in vacuum, treated as a purely geometric object that is free of matter yet carries a gravitational mass parameter \(M\). Building on the polynomial, \(g^2\)-regularized form of the Einstein equations, we use the Klinkhamer metric as a concrete example of such a matter‑free field source and recall that it reproduces the Schwarzschild exterior on each sheet while remaining an exact vacuum solution, including at the degenerate throat. We then formulate an extended equivalence principle which postulates that this geometric mass plays the role of both gravitational and inertial mass for the wormhole throat, viewed as the single dynamical degree of freedom. Applying this extended principle reduces the field‑theoretic problem of radial wormhole dynamics to the motion of a massive test particle in a Schwarzschild spacetime with mass \(M\), for which the collapse trajectory can be obtained analytically. We prove that any bound state of a traversable Klinkhamer wormhole inevitably collapses into a non‑traversable Einstein–Rosen wormhole, and we estimate the collapse time, showing that the traversable phase, although non‑stationary, can be a long‑lived matter‑free configuration.

 . 
\end{abstract}




\end{frontmatter}

Gravitational collapse is usually discussed in terms of self-gravitating matter configurations, such as pressureless dust or relativistic fluids, whose dynamics follow from Einstein’s equations supplemented by appropriate matter models. In this setting, the equivalence principle enters implicitly through the geodesic motion of test particles, but it does not by itself determine the full collapse dynamics. In the present work we investigate a complementary, purely gravitational scenario in which the collapsing object is not a matter distribution but a purely geometric, matter‑free wormhole throat endowed with mass \(M\).  This setting allows us to formulate and explore an extended version of the equivalence principle for matter-free sources of the gravitational field and to show that, in this case, the principle alone suffices to fix the radial collapse dynamics.

\section{Introduction}

In the pioneering work of Einstein and Rosen \citep{einstein}, it was shown that the source of the gravitational field, in addition to matter, is also curved space (i.e., the gravitational field itself). In this work, they constructed the first wormhole metric—the Einstein-Rosen bridge, connecting two sheets of the two-sheeted space. The Einstein-Rosen metric was \textit{degenerate}, since its determinant \(g\) vanishes on the spherical boundary that divides the two-sheeted space into two sheets. 

This circumstance made it impossible to describe the \textit{degenerate} metric within the framework of standard Einstein equations, since the determinant \(g\) appears as a denominator in Christoffel symbols—and hence in the components of the Riemann curvature. Einstein and Rosen solved this problem by multiplying the standard Einstein equations by \(g^2\). After this regularization, the Einstein equations become polynomial in the components of the metric tensor and their derivatives, allowing \textit{degenerate} metrics to be included in the description. Subsequently, the regularized form of the Einstein equations was called polynomial and was formalized within the framework of the Einstein-Hilbert variational approach \citep{peres,katanaev}.

Since the \textit{degenerate} Einstein-Rosen metric describes an empty and static curved space, the corresponding wormhole is an example of a matter-free stationary source of a gravitational field. In their original paper on the Einstein-Rosen bridge, its authors pointed out that the degenerate, geodesically complete solution they constructed of the regularized Einstein equations itself "describes the field-producing mass without requiring the introduction of any new field quantities" \citep{einstein}, Section 1.

For a long time, interest in wormholes remained purely academic. However, the work of Morris fnd Thorne \citep{MT} initiated the rapid development of concepts in this area \citep{visser}. The main goal of this research was to study the conditions for the existence and stability of a traversable wormhole. The starting point for this direction was Thorne's "engineering" approach, which first takes a suitable wormhole metric and then determines the required  matter distribution. A necessary element of this approach is the well-definiteness of the Riemann curvature, which is only possible with a \textit{non-degenerate} metric. Therefore, \textit{non-degenerate }wormhole states have long been (and remain) the primary object of study. The main obstacle in this direction was the inevitability of NEC violations \citep{hochberg}, which requires the presence of exotic matter to stabilize a traversable wormhole. Note that the proof of this constraint is entirely based on the applicability of standard Einstein equations and, consequently, on the non-degeneracy of the wormhole metric. 

Recent works by Klinkhamer \citep{klinkhamer1,klinkhamer2,klinkhamer3} and Wang\citep{wang} suggest a return to \textit{degenerate} wormholes, where it all began. The Klinkhamer wormhole metric is a direct generalization of the Einstein-Rosen metric. The generalization is that the radius \(b\) of a Klinkhamer wormhole can be larger than the gravitational radius, which ensures traversability. Constructed in complete analogy with the Einstein-Rosen metric, the Klinkhamer metric is also \textit{degenerate}. Therefore, it should also be described not by the standard Einstein equations but by regularized equations—for the same reason that Einstein and Rosen did so in 1935. Analogously to the Einstein–Rosen wormhole, the Klinkhamer wormhole constitutes a vacuum (purely geometric) solution that possesses an effective gravitational mass \(M\), without the need to invoke any additional matter sources.

This naturally raises the question: can such an object be stationary? The reason for this question is purely physical: as a "self-gravitating" object, the Klinkhamer wormhole should exhibit a tendency to contract, i.e. to gravitational collapse. This allows us to consider the dynamics and gravitational collapse of a \textit{degenerate} wormhole within a specific and well-controlled model. This problem is interesting both in itself and in light of a recent criticism of the Klinkhamer wormhole \citep{feng}, which points to the lack of uniqueness of the solution to the Cauchy problem in this class of metrics. 

This paper examines the dynamics of a degenerate wormhole using the example of the degenerate Klinkhamer wormhole. Specifically, the process of its gravitational collapse will be considered.

Investigating the dynamical properties of the Klinkhamer wormhole necessitates an appropriate generalization of its original stationary metric to a broader class of nonstationary metrics, in such a way that the stationary configuration arises as a particular case. Describing the dynamics of a \textit{degenerate} wormhole can be pursued through three possible approaches:

1) by employing \textit{exclusively} the regularized Einstein field equations themselves;

2) by formulating and analyzing a \textit{reduced action}, from which one derives the corresponding equations of motion for a suitably reduced set of dynamical variables;

3) by constructing a \textit{reduced description} grounded not on an action principle but rather on the \textit{equivalence principle}, which likewise leads to a system of equations governing a reduced set of variables.

Let us evaluate these three possibilities sequentially.

1) Restricting the analysis \textit{solely} to the regularized Einstein field equations is insufficient and reveals a fundamental limitation of this approach. The underlying reason is that the local metric \(g_{\mu\nu}\) by itself does not furnish a complete characterization of the wormhole. A full description is provided instead by the pair \((\mathscr{M}, g_{\mu\nu})\), which, in addition to the local metric \(g_{\mu\nu}\), incorporates the global topological properties of the manifold \(\mathscr{M}\). As demonstrated in \citep{dimaschko}, once this fact is taken into account, the application of the principle of least action yields, in a unified manner, both the Einstein field equations for \(g_{\mu\nu}\) and \textit{separate} evolution equations for the topological characteristics of the manifold \(\mathscr{M}\). It is precisely this circumstance that causes the aforementioned issue raised by Feng .

Specifically, in the context of the radial collapse of a degenerate wormhole, the sole topologically relevant degree of freedom is the time-dependent throat radius \(b(t)\). The Einstein field equations are \textit{insensitive} to the detailed temporal profile of \(b(t)\); they impose only that the spacetime metric reduces to the Schwarzschild form in the exterior domain \(r > b(t)\). As a result, the regularized Einstein equations, considered in isolation, are \textit{insufficient} to capture the complete dynamical evolution of a degenerate wormhole.

2) Employing a \textit{reduced action} would be a natural strategy in this context; however, it results in a highly intricate mathematical problem. The underlying reason is that the Einstein–Hilbert action functional, based on the scalar curvature, becomes ill-defined when the metric is degenerate. Consequently, it is necessary to resort to a polynomial action \citep{peres, katanaev}, which possesses a substantially more involved structure.

The formalism based on \textit{reduced variables} is also well established within the thin-shell model, which arises as the zero-thickness limit of the Thorne framework as the layer thickness approaches zero \cite{visser}. In this formulation, however, the spacetime metric remains nondegenerate and matter fields are present, thereby placing the Klinkhamer wormhole solution—characterized by a degenerate metric in the absence of matter—clearly outside the scope of this model.

3) Owing to its spherical symmetry, the degenerate Klinkhamer wormhole possesses a \textit{single} dynamical degree of freedom, namely the wormhole radius \(b\). Under these circumstances, the field dynamics effectively reduce to a one-dimensional problem, rendering it natural to invoke the \textit{equivalence principle}. This principle is conventionally formulated in the context of point-particle dynamics and, to date, has not been systematically employed to construct the dynamics of field configurations. The Klinkhamer wormhole thus provides the first known example of a spacetime metric that permits the direct application of the equivalence principle in this manner, thereby mapping a field-theoretic problem onto an equivalent single-particle problem.

This third approach is the one adopted in the present work. Its objective is to formulate the dynamical properties of a degenerate Klinkhamer wormhole on the basis of the equivalence principle. The proposed framework is predicated on extending the equivalence principle to field-theoretic (geometric) configurations that are free of matter sources. In the particular case of a degenerate wormhole, the central geometric object of interest is the wormhole throat.

The paper is organized as follows. In Sec. 2, we consider the regularized Einstein equations and the Klinkhamer metric as one possible stationary solution of this system of equations. In Sec. 3, we generalize the stationary Klinkhamer metric to a general non-stationary solution of the regularized equations, describing arbitrary wormhole dynamics within the framework of spherical symmetry. In Sec. 4, we extend the equivalence principle to matter-free objects, an example of which is a \textit{degenerate} wormhole. In Sec. 5, we directly apply the  equivalence principle, which allows us to reduce the dynamics of a spherically symmetric wormhole to the dynamics of a single test particle in a Schwarzschild gravitational field. In Sec. 6,  we present a well-known solution to the problem of the radial fall of a test particle in a Schwarzschild gravitational field.  In Sec. 7, this allows us to obtain a unique solution to the problem of the dynamics of a \textit{degenerate} spherical wormhole under given initial conditions, in particular, a complete solution to the gravitational collapse problem. The collapse results in the transformation of the Klinkhamer wormhole into an Einstein-Rosen wormhole. In the same section, we also examine the consistency between the equations of motion derived from the equivalence principle and the dynamical equations obtained in the Newtonian limit. Finally, in Sec. 8, we summarize and draw conclusions. 

\section{Stationary metric}

We begin by presenting the Klinkhamer metric describing a \textit{degenerate} spherical wormhole in a vacuum. We choose the Schwarzschild metric as the initial one\footnote{The relativistic system of units is used, in which the speed of light \(c\) and the gravitational constant \(G\) are equal to 1, as well as the usual spherical coordinates \((r, \theta, \varphi)\),  \( \; d\Omega^2= d\theta^2+sin^2 \theta   d\varphi^2 \) is an element of solid angle, \(t\) is time, \(M\) is the mass of the source of the gravitational field.}
\begin{equation}
    ds^{2} = -\left(1 - \frac{2M}{r}\right) dt^{2}
    + \left(1 - \frac{2M}{r}\right)^{-1} dr^{2}
    + r^{2} d\Omega^{2}.
\end{equation} It is a solution to the standard Einstein equations in a vacuum\footnote{The standard notation is adopted: \(g_{\mu\nu}\) is the metric tensor, \(R_{\mu\nu}\) and \(R\) are the Ricci tensor and its trace.}
\begin{equation}
    R_{\mu \nu} - \frac{1}{2} g_{\mu \nu} R = 0,
\end{equation}defined in a one-sheeted space with a non-negative radial coordinate \(r\ge 0\). We apply the topological dressing procedure of the two-sheeted coordinate transformation \citep{dimaschko} to the initial metric (1). To do this, we choose the following coordinate transformation
\begin{equation}
    r = \sqrt{l^2 + b^2},
\end{equation}which transforms the original one-sheeted space into the two-sheeted space of the wormhole. The new variable \(l\) covers all values from \(-\infty\) to \(+\infty\), with positive values \(l>0\) corresponding to the first sheet, negative values \(l<0\) to the second sheet of the two-sheeted space, and the value \(l=0\) to the transition surface between them (i.e., the spherical throat \(r=b\)). The constant parameter \(b \ge 2M\) represents the radius of the throat. The topological dressing transforms the original solution (1) of the standard Einstein equations (2) into a \textit{degenerate} Klinkhamer metric \citep{klinkhamer1}
\begin{equation}
   \begin{split}
 ds^{2}
    = -\left(1 - \frac{2M}{r}\right) dt^{2}
    + \left(1 - \frac{2M}{r}\right)^{-1}
      \frac{l^{2}}{r^{2} } dl^{2}
    + r^{2} d\Omega^{2}.
   \end{split}
\end{equation} This new metric, defined on a two-sheeted space, is a solution of the \textit{regularized} Einstein equations in vacuum
\begin{equation}
    g^2 \left( R_{\mu \nu} - \frac{1}{2} g_{\mu \nu} R \right) = 0
\end{equation} Despite the absence of matter, the \textit{degenerate} wormhole (4) itself is the source of a gravitational field, which outside the throat is described by the Schwarzschild metric. To verify this, it is sufficient to convert the metric (4) to ordinary radial variables using the inverse of transformation (3):
\begin{equation}
    l = \pm \sqrt{r^2 - b^2} \quad (r \geq b),
\end{equation} Here, the choice of the “+” or “-” sign corresponds to the first or second sheet, respectively. After this transition to the usual \(r\)-coordinate, metric (4) returns to the form of the Schwarzschild metric (1), but with the constraint \(r\ge b\):
\begin{equation}
    ds^{2} = -\left(1 - \frac{2M}{r}\right) dt^{2}
    + \left(1 - \frac{2M}{r}\right)^{-1} dr^{2}
    + r^{2} d\Omega^{2},
    \quad (r \geq b).
\end{equation}
This means that in the entire domain of definition \(r\ge b\) of the metric (7), it coincides with the original Schwarzschild metric (1) – i.e., the matter-free throat creates on each of the two sheets the same gravitational field as a spherically symmetric distribution of matter with a total mass \(M\).

\section{Non-stationary metric}
Thus, the Klinkhamer metric (4) describes a massive wormhole with mass \(M\). This should lead to the instability of such a wormhole with respect to gravitational collapse, during which the radius of the throat \(b\) decreases.
 
We will construct a non-stationary solution describing this process. To do this, in the original Schwarzschild metric (1), we perform a double coordinate substitution \((t,r)\mapsto(b,l)\) as follows:
\begin{equation}
    r = \sqrt{l^2 + b^2}, \quad t = t(b).
\end{equation}
Here \(t(b)\) is a function that must be determined later. This substitution, in addition to the two-sheeted transformation of the radial coordinate, analogous to transformation (3) , also transforms the radius \(b\) of the wormhole into a new time variable. 

The core feature of this transformation is that the conventional time coordinate \(t\) is reparameterized in terms of the quantity \(b\), representing the instantaneous radius of the wormhole. This procedure replaces \(t\) with a new time coordinate \(b\), which possesses a clear geometric interpretation: an observer assigns time values according to the current radius of the throat \(b\). \textit{Consequently, the collapsing wormhole is effectively reinterpreted as a geometric clock, with its dynamical geometry serving as a measure of time.}

In the new variables, the non-stationary metric has the form
\begin{equation}
    ds^{2}
    = -\left(1 - \frac{2M}{r}\right)\dot{t}^{2}\,db^{2}
    + \left(1 - \frac{2M}{r}\right)^{-1}
      \frac{\left(l\,dl + b\,db\right)^{2}}{r^{2} }
    + r^{2}d\Omega^{2},
\end{equation}where \(\dot{t}=dt/db\). This metric describes a wormhole of variable radius, which is given by the function \(b(t)\), the inverse of \(t(b)\). Obtained from the Schwarzschild metric (1) by a coordinate transformation, this metric, like the stationary Klinkhamer metric (4), is a solution to the regularized Einstein equations (5). This follows directly from the covariance of Einstein's equations and is, perhaps surprisingly, completely independent  of the choice of the function \(b(t)\).

This independence from the function \(b(t)\) arises because Einstein's equations, which govern only the local metric, are not sufficient to fully describe the dynamics of space with a \textit{changing boundary}\footnote{The recognition that the wormhole throat constitutes an independent topological variable has precedent in the thin-shell wormhole literature \citep{visser2,visser}, where the throat radius enters as the configuration variable in a Wheeler-De Witt quantization. The dynamical or statistical treatment of the topological variables has long been considered in discussions of topology change \citep{geroh,sorkin,horowitz,deborde}, wormhole formation \citep{hawking,visser,koga}, and topological quantum field theory \citep{atiyah,witten}. More recently, \citep{borissova} derived an effective action for throat dynamics from first principles, showing that the throat radius is 'a single degree of freedom' governed by its own equation of motion derived from the Israel junction formalism. \citep{dimaschko} extended this concept to \textit{degenerate} wormholes, where the standard Einstein equations are insufficient. It was also shown that a direct application of the principle of least action allows one to include topological degrees of freedom in the analysis.}.   As shown in \citep{dimaschko}, for such a complete description, the principle of least action must be directly used. On the one hand, this principle provides Einstein's equations themselves, which determine the local geometry of the space. On the other hand, this same principle determines the global topology of the space and, in particular, the dynamics of the wormhole boundary.

A similar approach could be taken to determine the dependence \(b(t)\) that describes the collapse of a traversable wormhole. However, here we will do this more simply, by directly applying the equivalence principle.

\section{ The equivalence principle}
In the analysis of the dynamics of a \textit{degenerate} wormhole subjected to its self-generated gravitational field, it is essential to identify and characterize the source of this field. This issue can be illustrated by considering the specific example of the Klinkhamer wormhole.

To proceed, we compare the Schwarzschild metric (1) with the Klinkhamer metric (7), expressed in the same radial coordinate \(r\). Both metrics involve the same mass parameter \(M\), but its physical interpretation differs in the two cases. In the Schwarzschild metric, \(M\) represents the mass of a spherically symmetric distribution of matter in three-dimensional space. By contrast, the Klinkhamer metric is an exact solution of the regularized Einstein equations (5) on the entire two-sheeted manifold, where no matter is present. In this latter case, the source of the gravitational field is not matter in the usual sense, but rather the spherical two-surface  associated with the two-sheeted geometry—specifically, its spherical throat at \(r = b\). Owing to the symmetry of the two-sheet metric (4) under the exchange of the two sheets (equivalently, the change of sign of the two-sheet coordinate \(l\)), the resulting gravitational field is identical on both sheets.

Thus, the wormhole can be regarded as an object endowed with a gravitational mass \(M\). A natural question then arises: does this object also possess inertia? Without addressing this issue, it is not possible to formulate a consistent dynamical description of a wormhole. To resolve this, we invoke the equivalence principle, which postulates 

 \vspace{12pt}
 
  \fbox{\parbox{0.44\textwidth}{
“the equality (in suitable units) of the inertial and gravitational masses, regardless of the nature of the \textit{body}” \citep{MTW}. 
}}

 \vspace{12pt}

In conventional usage, the term “\textit{body}” refers to a material distribution occupying a three-dimensional spatial region. In the wormhole case, however, the situation is fundamentally different: we are dealing with a compact, gravitating two-dimensional surface that is devoid of matter. Accordingly, the standard formulation of the equivalence principle must be generalized to assert 

 \vspace{12pt}
 
  \fbox{\parbox{0.44\textwidth}{
the equality (in suitable units) of the inertial and gravitational masses, regardless of the nature \textit{and dimensionality of the object}. 
}}

 \vspace{12pt}

What criteria determine the admissibility of such an extension? A necessary condition for its acceptability is the absence of internal inconsistencies, as well as the successful incorporation of its implications into an already established and empirically tested system of theoretical concepts and operationally defined, physically measurable quantities.

In Section 7 of this paper, we show that in the Newtonian limit, the extended formulation of the equivalence principle precisely corresponds to the principles of ordinary classical mechanics extended to a two-sheeted space. In this sense, it passes the test of the correspondence principle  and allows for a natural extension to general relativity. Therefore, in our further consideration of the radial dynamics of the Klinkhamer wormhole, we will always proceed from this extended interpretation of the equivalence principle. 

\section{Gravitational collapse and the equivalence principle}

\textit{In this section, we show that a direct application of the equivalence principle reduces the field problem of the radial dynamics of a spherically symmetric wormhole (here, the Klinkhamer wormhole) to a single-particle problem of the motion of a test particle in a given gravitational field.}

Note first that the use of new (\(b,l)\)  coordinates instead of the usual variables \((t,r)\)  according to substitution (8) is nothing more than a basis for representing the non-stationary solution in the simple form (9). This representation is convenient for verifying the mathematical consistency of the solution with Einstein's equations. On the other hand, for the physical analysis and interpretation of the solution, it is more convenient to use the usual variables \((t,r)\), which is what we will do next. In these ordinary coordinates, the wormhole surface appears simply as a sphere of variable radius \(r=b(t)\), and the non-stationary metric (9) again has the form of a Schwarzschild metric with the constraint
\begin{equation}
    ds^{2} = -\left(1 - \frac{2M}{r}\right) dt^{2}
    + \left(1 - \frac{2M}{r}\right)^{-1} dr^{2}
    + r^{2} d\Omega^{2},
    \quad (r \geq b(t)).
\end{equation}
Let the radius of a collapsing wormhole decrease according to the law \(b(t)\) from an initial value \(b(0)>2M\). Consider the free motion of a test particle of mass \(m\), located at the initial time \(t=0\) near the wormhole surface, at a point with radial coordinate \(r(0)=b(0)+0\). The test particle is assumed to be initially at rest relative to the surface of the wormhole.

By virtue of the equivalence principle, the position of the test particle relative to the wormhole surface should remain unchanged during the collapse process (the "falling elevator effect"). This means that \(b(t)\) must be identical to the law of radial motion of a particle \(r(t)\) in a given gravitational field of the wormhole (see Fig. 1).  This will allow us to reduce the field problem of throat motion to a single-particle problem of motion of a test particle in a given field simply by replacing \(r(t)\) with \(b(t)\) in the equation of particle motion.

Thus, the problem of wormhole collapse is reduced to the problem of free fall of a test particle located near the surface of a collapsing wormhole.\footnote{In this case, the possibility of applying the equivalence principle is due to the absence of non-gravitational forces. In \citep{dimaschko}, a charged wormhole was considered against the background of an electrovacuum. In this case, the equivalence principle was inapplicable due to the presence of non-gravitational forces. Therefore, to determine the equilibrium radius of the wormhole, a direct application of the principle of least action was required. 
}  

\begin{figure} [h]
 \centering
 \includegraphics[scale=0.65]{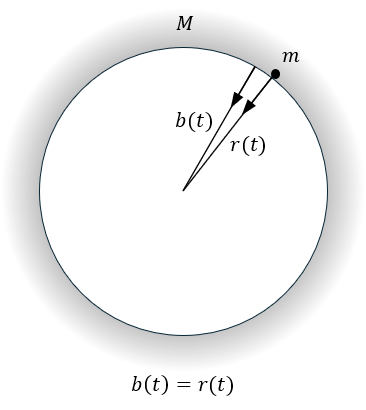}
  \caption{By virtue of the equivalence principle, the collapse of a \textit{degenerate} wormhole of mass \(M\) is synchronous with the free fall of a test particle of mass \(m\) located near the surface of the wormhole.}
\end{figure}

\section{Free fall of a test particle}
The free fall of a test particle of mass \(m\) can be described by the Hamilton-Jacobi equation for a single-particle action \(S\):
\begin{equation}
    g^{\mu\nu} \frac{\partial S}{\partial x^{\mu}} \frac{\partial S}{\partial x^{\nu}} = m^{2}.
\end{equation} In the case of Schwarzschild geometry (1) and radial motion of the test particle of interest here, it has the form
\begin{equation}
    \left(1 - \frac{2M}{r}\right)^{-1} \left(\frac{\partial S}{\partial t}\right)^{2}
    - \left(1 - \frac{2M}{r}\right) \left(\frac{\partial S}{\partial r}\right)^{2}
    = m^{2}.
\end{equation} We seek the solution of this equation in the form
\begin{equation}
    S = -\varepsilon t + S_r(r),
\end{equation} where \(\varepsilon\) is the total energy of the particle in the gravitational field. This yields an integral expression for \(S_r (r)\):
\begin{equation}
    S_r(r) = \pm \int \left[ \left(1 - \frac{2M}{r}\right)^{-2} \varepsilon^2 - \left(1 - \frac{2M}{r}\right)^{-1} m^2 \right]^{1/2} dr.
\end{equation} Here, the choice of sign determines the direction of motion of the test particle relative to the radial axis. The condition \(\partial S / \partial \varepsilon = 0\) yields an integral dependence of \(t\) on \(r\)
\begin{equation}
    t = \pm \frac{\varepsilon}{m} \int \left(1 - \frac{2M}{r}\right)^{-1} 
    \left[\left(\frac{\varepsilon}{m}\right)^2 - \left(1 - \frac{2M}{r}\right)\right]^{-1/2} dr,
\end{equation}
which can be transformed into a differential equation
\begin{equation}
    \frac{dr}{dt} = \pm \left(1 - \frac{2M}{r}\right) \left[1 - \left(\frac{m}{\varepsilon}\right)^2 \left(1 - \frac{2M}{r}\right)\right]^{1/2}.
\end{equation}This equation determines the relationship between the radial velocity of a test particle and its radial coordinate. We further use this relationship to construct a phase portrait of a \textit{degenerate} wormhole, describing, in particular, its gravitational collapse.

\section{Radial dynamics and self-energy of the wormhole}
We have demonstrated that, by virtue of the equivalence principle, the time dependence of the wormhole radius \(b(t)\) must coincide with the time dependence \(r(t)\) describing a freely falling test particle. Consequently, the functions \(b(t)\) and \(r(t)\) are required to satisfy the same differential equation, namely the one given in the form of equation (16) for \(r(t)\). It follows that the evolution equation for the wormhole radius \(b(t)\) can be obtained directly from equation (16) by substituting \(b(t)\) for \(r(t)\):
\begin{equation}
    \frac{db}{dt} = \pm \left(1 - \frac{2M}{b}\right) \left[1 - \left(\frac{M}{\mathcal{E}}\right)^2 \left(1 - \frac{2M}{b}\right)\right]^{1/2}.
\end{equation}Here, the minus sign corresponds to the case of gravitational collapse (a decrease in the radius \(b\)), and the plus sign corresponds to the opposite case of inertial expansion (an increase in the radius \(b\)). Simultaneously, we replace the mass \(m\) of the test particle with the mass \(M\) of the wormhole and the total energy \(\varepsilon\) of the test particle in a given gravitational field with the "self-energy" \(\mathcal{E}\) of the wormhole. 

In fact, the quantity \(\mathcal{E}/M\), which appears in the new differential equation (17), inherits from the ratio \(\varepsilon /m\), which enters the original differential equation (16), only its \textit{kinematic} meaning—the meaning of the first integral of motion. This meaning, in complete analogy with (16), consists of nothing more than the conservation of a certain combination of the radius \(b\) of the wormhole and the rate \(db/dt\) of change of this radius:

\begin{equation}
    \left(1 - \frac{2M}{b}\right)^{-1} \left[ 1 - \left(1 - \frac{2M}{b}\right)^{-1} \left(\frac{db}{dt}\right)^{2} \right] = \text{const} = \left(\frac{M}{\mathcal{E}}\right)^{2}.
\end{equation}
To clarify the independent dynamic meaning of the "self-energy" \(\mathcal{E}\), we will proceed in the same way as is customary to clarify the meaning of the formal parameter \(M\) in the Schwarzschild solution: namely, we will move to the Newtonian limit. In the context of the equation of motion (18), this implies a lower-order approximation in the small parameter \(2M/b\), within which equation (18) takes the form
\begin{equation}
    \mathcal{E} = M + \frac{M}{2} \left( \frac{db}{dt} \right)^{2} - \frac{M^{2}}{b}.
\end{equation}
Let us compare this expression with the result that follows directly from Newton's theory of the gravitational field for the same spherical wormhole of mass \(M \) and radius \(b\). This requires calculating the corresponding energy \(U\) of the gravitational field, which in the Newtonian approximation is a locally well-determined quantity:
\begin{equation}
    U = -\frac{1}{8\pi} \int \mathfrak{g}^2 \, dV.
\end{equation}Here, \(\mathfrak{g}\) is the local acceleration of gravity, which in the spherically symmetric case has an absolute value
\begin{equation}
    g = \frac{M}{r^2}, \quad (r \geq b),
\end{equation}and the integration extends to the entire two-sheeted space. This corresponds to a \textit{double} integration of expression (20) over the region \(r\ge b\) of the ordinary one-sheeted space and leads to the following expression for the total gravitational energy:
\begin{equation}
    U(b) = -\frac{2}{8\pi} \int_{r \geq b} g^2 \, dV = -\frac{M^2}{b}
    \label{eq:placeholder_label}.
\end{equation}The factor 2 in front of the integral in this expression appears precisely because of the two-sheeted nature of space and the double integration over the region \(r\ge b\) . After this, the Newtonian wormhole Lagrangian, written in the usual form,
\begin{equation}
    L\!\left(b,\frac{db}{dt}\right) = \frac{M}{2}\left(\frac{db}{dt}\right)^{2} - U(b),
\end{equation}immediately leads to the conservation of the total energy of the wormhole, which includes the energy of the field (22):
\begin{equation}
    \frac{M}{2} \left( \frac{db}{dt} \right)^{2} - \frac{M^{2}}{b} = \text{const.} 
\end{equation}
By denoting the constant on the right-hand side as \(E-M\), we arrive at the law of conservation of the total energy \(E\)  of the wormhole:
\begin{equation}
    E = M + \frac{M}{2} \left( \frac{db}{dt} \right)^{2} - \frac{M^{2}}{b}.
\end{equation}This expression for the total energy \(E\) coincides with the previous expression (19) for the formal parameter \(\mathcal{E}\).

Thus, in the Newtonian limit, the formal conserved quantity \(\mathcal{E}\) arising in the equation of motion (18) naturally reproduces the wormhole's total self-energy \(E\), which includes the field energy. This provides grounds for maintaining the same interpretation in the relativistic case, where the gravitational field energy is no longer a\textit{ locally}, but rather a \textit{globally} defined quantity. For this reason, we shall henceforth refer to the corresponding integral of motion \(\mathcal{E}\) as the \textit{wormhole's self-energy}, without any further qualifications.

We have thus demonstrated that, in the Newtonian limit, the \textit{extended formulation} of the equivalence principle we adopted is completely consistent with ordinary classical mechanics — both lead to identical equations of motion (19) and (25). It is precisely the correspondence principle that establishes the consistency of this extended formulation with classical mechanics, which is initially based on our daily experience. A rigorous derivation of the extended equivalence principle from the polynomial action is a much more difficult task and is left for future research. 

Equation (17) completely determines the phase portrait of the Klinkhamer wormhole. This phase portrait in coordinates \((b,db/dt)\) is shown below in Fig. 2. The separatrix here is the phase trajectory for which \(E=M\). In the phase plane, it bounds the region within which \(E<M\), and therefore the radius of the wormhole can vary only within finite limits (bound states). 

Overall, the resulting phase portrait describes both the gravitational contraction of the wormhole (the lower part of the phase plane – below the separatrix) and its inertial expansion (the upper part of the phase plane – above the separatrix). In the first case, all phase trajectories converge at the point \((2M,0)\), corresponding to a non-traversable Einstein-Rosen wormhole. In the second case, the asymptotic limit of the process is the uniform inertial expansion of a traversable wormhole.

\begin{figure} [t]
 \centering
 \includegraphics[scale=0.42]{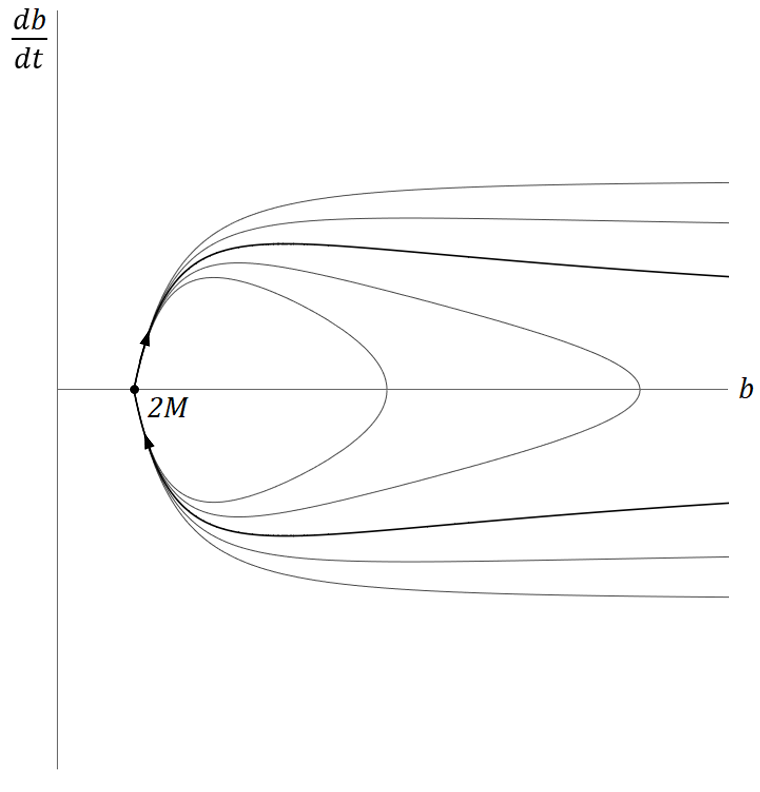}
  \caption{Phase portrait of the Klinkhamer wormhole dynamics. The lower part of the phase portrait describes the gravitational contraction of the wormhole, and the upper part describes its inertial expansion. The separatrix, shown by the thick line, bounds the region \(E<M\), where the wormhole radius can vary only within finite limits. In this region, as in the region of gravitational contraction, the evolution of the Klinkhamer wormhole always ends with gravitational collapse into an Einstein-Rosen wormhole. This configuration corresponds to the stationary point \( (2M,0)\), at which the wormhole throat radius coincides with the Schwarzschild radius \(2M\).}
\end{figure}

In the intermediate region, bounded by the separatrix, the radius of the wormhole varies within finite limits and culminates in a gravitational collapse at the point \((2M,0)\).

Note that here the limit \(b \rightarrow 2M\) does not produce the usual Schwarzschild interior region \(0<r<2M\), but rather the Einstein–Rosen bridge connecting two copies of the exterior Schwarzschild region \(r\ge2M\) via the degenerate throat at \(r=2M\). In the Einstein-Rosen-Horowitz-Klinkhamer polynomial formulation, this configuration is geodesically complete and free of curvature singularities on both sheets; the singular region \(r=0\) of the standard maximally extended Schwarzschild solution is simply absent from the manifold.

The process of gravitational collapse of the Klinkhamer wormhole into a non-traversable Einstein-Rosen wormhole, corresponding to trajectories in the lower half of the phase portrait, can be conveniently represented as the evolution of a Flamm paraboloid describing the wormhole. This evolution is shown in Fig. 3.

\begin{figure} [h]
 \centering
 \includegraphics[scale=0.42]{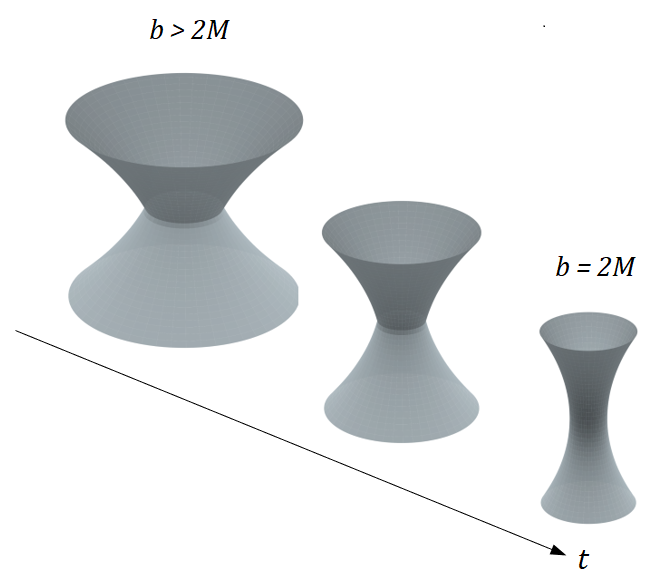}
  \caption{During gravitational collapse, the traversable Klinkhamer wormhole \((b>2M)\) transforms into the non-traversable Einstein-Rosen wormhole \((b=2M)\). The Flamm paraboloid has a kink at the throat, which disappears as a result of collapse (i.e., as \(b \rightarrow 2M\)). The three Flamm paraboloids shown are assumed to have identical values of the mass parameter \(M=1\) and successively decreasing values of the throat radius: \(b=6\), \(b=4\), and \(b=2\).}
\end{figure}
As shown in Appendix A, the Flamm paraboloid of the Klinkhamer metric has a kink at the throat. During gravitational collapse, the kink becomes less pronounced as the throat radius decreases until it completely disappears in the Einstein-Rosen wormhole limit (see Fig. 3). The geometric interpretation of this kink is that, along the dynamically evolving throat surface \(r = b(t)\), the induced spatial metric remains degenerate. The kink is eliminated only in the final configuration \(b = 2M\), when the throat merges with the event horizon. In this state, spatial radial degeneracy is precisely compensated by the relativistic radial factor \((1 - 2M/r)^{-1}\).

The physical picture that emerges is intuitive: a traversable \textit{degenerate} wormhole, lacking any matter to support it, is gravitationally unstable. Under its own gravitational attraction, the throat contracts until it reaches the minimum radius \(b=2M\), at which point the wormhole becomes an Einstein-Rosen bridge—the 'maximally collapsed' configuration of a \textit{degenerate} wormhole. This final state is stable against further contraction since \(b=2M\) corresponds to the Schwarzschild horizon, where the collapse velocity vanishes.

By ‘stable final state’ we mean that the collapsing degenerate wormhole asymptotically approaches this static Einstein–Rosen configuration and cannot evolve further. No further contraction beyond \(b=2M\)  is possible because the manifold by construction has no region with \(r<2M\).

To conclude this section, let us dwell on the physical interpretation of the phase portrait shown in Fig. 2 in more detail  This phase portrait describes three possible regimes:

1) collapse from an unbound state with \(E>M\) (phase trajectories below the separatrix);

2) collapse from a bound state with \(E<M\) (phase trajectories inside the separatrix);

3) the regime of unlimited expansion with \(E>M\) (phase trajectories above the separatrix).

In each of these three regimes, the Klinkhamer wormhole, although non-stationary, nevertheless remains traversable for a certain time. This time depends on the initial conditions and can be quite significant. For example, the collapse from a resting state with an initial radius \(b(0) \equiv  b_0 \gg 2M\) takes a time \(\tau\) equal to the fall time of a test particle from a distance \(b_0\) in the gravitational field of a classical point source of mass \(M\). This time is \citep{landau1}, Eq.(103.17):
\begin{equation}
    \tau = \frac{\pi}{2} b_0 \sqrt{\frac{b_0}{GM}}. 
\end{equation}
For example, for the Klinkhamer wormhole with radius \(b_0=10\) m and mass \(M=10^3 \) kg, the gravitational collapse time is \(\tau =1.9 \cdot 10^5 \) s, i.e., approximately 2 days. Of course, the interpretive value of this estimate is limited, as the question of the mechanism by which the initial state of a wormhole of radius \(b_0\) is formed remains unanswered. However, this estimate shows that the Klinkhamer wormhole, although non-stationary, is a relatively long-lived object. It does not disappear "at the speed of light." 

While our present analysis focuses on the collapse dynamics, a full study of the formation of degenerate metric states would require linking them to specific physical processes involving matter. To study the conditions for the formation of a state with a \textit{degenerate} metric, this metric must be linked to a specific physical process involving matter. For example, in astrophysics, this could  involve the evolution of a neutron star, the loss of stability of a stellar state, the formation of a supernova, etc. To account for the possibility of states with a \textit{degenerate} metric, these processes must be considered within the framework of \textit{regularized} Einstein equations — but not the standard Einstein equations, as is currently the case.

\section{Conclusions}

Thus, using the Klinkhamer wormhole as an example, we have described the radial dynamics of a \textit{degenerate}, spherically symmetric wormhole. A special case of these dynamics is gravitational collapse, transforming the Klinkhamer wormhole into a non-traversable Einstein-Rosen wormhole.

The result obtained here should be compared with the objection raised by Feng  \citep{feng}, who pointed out the apparent inconsistency of the non-stationary generalization of the Klinkhamer metric - and, consequently, of the Klinkhamer metric itself. Feng noted that the function \(\lambda (t)\) (our \(b(t)\)) is not constrained by the Einstein equations, which makes the initial-value problem ill-posed and seemingly  deprives the Klinkhamer metric of any physical meaning.

The resolution of the contradiction is the following:

1) the Einstein equations do indeed allow an arbitrary choice of the function \(b(t)\). Feng’s observation is correct in this respect, and this is fully confirmed in Section 3 of our work; 

2) however, \textit{not every} choice of \(b(t)\) describes the \textit{real} dynamics of the wormhole.

The reason is that the local metric \(g_{\mu \nu}\) alone does not provide a complete description of the wormhole. The complete description is given by the pair \((\mathscr{M},g_{\mu \nu})\), which, along with the local metric \(g_{\mu \nu}\), additionally includes the global characteristics of the manifold \(\mathscr{M}\). In \citep{dimaschko}, it is shown that, with this in mind, applying the principle of least action simultaneously yields both Einstein's equations for \(g_{\mu \nu}\) and separate equations for the topological characteristics of the manifold \(\mathscr{M}\). In the problem of the radial collapse of a wormhole, the only such characteristic is \(b(t)\), the radius of the wormhole. 

Due to spherical symmetry, the solution to this problem is significantly simplified by the possibility of applying the equivalence principle - \textit{extended }to account for the presence of matter-free sources of the gravitational field. Direct application of the equivalence principle reduces the radial collapse problem of the Klinkhamer wormhole to the well-known problem of a test particle falling in a Schwarzschild gravitational field. This uniquely determines the dynamical equation for \(b(t)\). As follows from the phase portrait in Fig. 2, the corresponding Cauchy problem has a unique solution for any initial conditions (\(b_0,\dot{b_0}\)).

Thus, the contradiction discovered by Feng is resolved.

The correspondence established in this work between the radial dynamics of the Klinkhamer wormhole and the fall of a test particle in a Schwarzschild gravitational field allows us to determine the time of gravitational collapse. Simple estimates show that the Klinkhamer wormhole, although non-stationary, is a relatively long-lived state, realizing the idea of a traversable wormhole.

The results presented in this work are derived from the application of the equivalence principle under the assumption of spherical symmetry. These  two factors enable us to reduce the original field-theoretic problem to an effective single-particle description, thereby allowing us to obtain the dynamics of a degenerate wormhole in explicit form. Besides spherical symmetry, a key prerequisite for  this approach is the absence of matter or, more generally, the predominance of gravitational interactions over other fundamental forces. 

Within this framework, an entire class of problems can be identified whose central focus is the dynamical behavior of degenerate wormholes. In contrast to the conventional case of \textit{nondegenerate} wormholes, \textit{degenerate} wormholes constitute substantially simpler objects from both a mathematical and a physical standpoint. Furthermore, their existence does not require the presence of exotic matter, which provides a strong motivation for further investigation along these lines.

As a subsequent step, we plan to consider the quantum dynamics of a degenerate wormhole modeled as a system with a single dynamical degree of freedom, namely the wormhole radius. The corresponding analysis has been completed and will be reported in a forthcoming publication.

\appendix

\section{Flamm's paraboloid of the Klinkhamer metric}

The general definition of a Flamm paraboloid for an arbitrary spherically symmetric metric follows from the local Pythagorean relation, which expresses the square of the differential of the proper radial coordinate \(dl^2\) through the sum of the squares of the differentials of the ordinary \(r\)-coordinate \(dr^2\) and the auxiliary \(z\)-coordinate \(dz^2\):
\begin{equation}
    dl^2 = dr^2 + dz^2
\end{equation}Taking into account (9), the square of the differential of the proper radial coordinate is
\begin{equation}
    dl^2 = \left(1 - \frac{2M}{r}\right)^{-1} dr^2,
\end{equation}which yields the differential equation
\begin{equation}
    dz = \pm \left[ \left(1 - \frac{2M}{r}\right)^{-1} - 1 \right]^{1/2} \, dr.
\end{equation}The solution, expressed as \(r(z)\), determines the shape of the Flamm paraboloid for the Klinkhamer metric in cylindrical coordinates \((r,\varphi ,z)\). For the unique determination of the solution \(r(z)\), differential Eq. (A.3) requires a boundary condition, which we choose in the form
\begin{equation}
    r(0) = b.
\end{equation}Its geometric meaning is that the throat, with radius b, corresponds to a cross-section of the paraboloid by the plane \(z=0\). From this condition, we obtain 
\begin{equation}
    r(z) = b + \frac{z^2}{8M} + \lvert z \rvert \sqrt{\frac{b}{2M} - 1}
\end{equation}The Flamm paraboloid of the Klinkhamer metric has a kink at the throat (that is, at \(z=0\)), which disappears at \(b=2M\).

\section*{Declaration of Competing Interest}

The author declares that he has no known competing financial interests or personal relationships that could influence the content of this paper.

\section*{Data availability}

No data was used for the research described in the article.

\bibliographystyle{elsarticle-harv} 
\bibliography{main}






\end{document}